\documentclass[grl]{agu2001}

\usepackage{graphicx}

\authorrunninghead{THOMAS ET AL.}

\titlerunninghead{1859 Solar Flare Effects}

\authoraddr{B. C. Thomas,
Department of Physics and Astronomy, Washburn University, 1700 SW College Ave.,
		 Topeka, KS 66621, USA.
(brian.thomas@washburn.edu)}

\begin{document}

\title{Modeling atmospheric effects of the September 1859 Solar Flare}

\author{B. C. Thomas}
\affil{Department of Physics and Astronomy, Washburn University, Topeka, Kansas, USA}

\author{C. H. Jackman}
\affil{Laboratory for Atmospheres, NASA Goddard Space Flight Center, Greenbelt, Maryland, USA}

\author{A. L. Melott}
\affil{University of Kansas, Department of Physics and Astronomy, Lawrence, Kansas, USA}

\begin{abstract} 
We have modeled atmospheric effects, especially ozone depletion, due to a solar proton event which probably accompanied the extreme magnetic storm of 1-2 September 1859.  We use an inferred proton fluence for this event as estimated from nitrate levels in Greenland ice cores.  We present results showing production of odd nitrogen compounds and their impact on ozone.  We also compute rainout of nitrate in our model and compare to values from ice core data.
\end{abstract}

\begin{article}

\section{Introduction}
Atmospheric ozone effects by solar proton events (SPEs) associated with solar eruptive events have been studied since the 1970's.  Ozone depletion occurs following the production of odd hydrogen- and nitrogen-oxides. Production of $\mathrm{HO_x}$ (e.g., H, OH, $\mathrm{HO_2}$) and $\mathrm{NO_y}$ (e.g., N, NO, $\mathrm{NO_2}$, $\mathrm{NO_3}$) and depletion of $\mathrm{O_3}$ by solar proton events has been studied through both satellite observations and computational modeling (e.g., \citet{cru75,he77,reag81,jm85,jack00,jack01}).  \citet{jm04} review much of the work in this area.

The solar flare of 1 September 1859 was one of the most intense white-light flares ever observed \citep{tsu03}.  The flare itself was observed independently by \citet{carr60} and \citet{hodg60} and lasted approximately 5 minutes.  The flare was followed about 17 hours later by a magnetic storm at the Earth which lasted about 2 hours \citep{carr60,tsu03}.  The storm was of such intensity that in the United States and Europe fires were started after arcing from induced currents in telegraph wires \citep{loom61}.  The storm was likely caused by energetic charged particles accelerated by one or several highly energetic coronal mass ejections (CME) from the Sun at the time of the flare.  The geomagentic activity associated with the accelerated particles lasted several days at least \citep{cb40}.  Studies of very energetic events associated with solar activity are important in understanding how such activity impacts various Earth-based systems.  An event as energetic as the 1859 one has not been modeled in this way before, and it may be that events of this magnitude and larger are not uncommon over the long term \citep{sch00,sm07}.

\section{Methods}  
Our modeling was performed using the Goddard Space Flight Center two-dimensional atmospheric model that has been used previously for modeling SPEs \citep{jack90,jack00,jack01,jack05a,jack05b}, as well as much higher energy events such as gamma-ray bursts \citep{thom05}.  We briefly describe the model here.  More detail on the version of the model used and its reliability for high fluence events is given by \citet{thom05} and the appendix therein.

The model's two spatial dimensions are altitude and latitude.  The latitude range is divided into 18 equal bands and extends from pole to pole.  The altitude range includes 58 evenly spaced logarithmic pressure levels (approximately 2 km spacing) from the ground to approximately 116 km.  The model computes 65 constituents with photochemical reactions, solar radiation variations, and transport (including winds and small scale mixing) as described by \citet{thom05} and \citet{flem99}.   
A photolytic source term is computed from a lookup table and used in calculations of photodissociation rates of atmospheric constituents by sunlight \citep{jack96}.

We have employed two versions of the atmospheric model.  One is intended for long term runs (many years) and includes all transport mechanisms (e.g., winds and diffusion); it has a time step of one day and computes daily averaged constituent values.  The second is used for short term runs (a few days) and calculates constituent values throughout the day and night, but does not include full transport.  This version has a time step of 225 seconds.

No direct measurements of the proton fluence are available from 1859, but an estimate of the fluence of this event based on measurements of nitrate enhancement in Greenland ice cores is given by \citet{mcc01a,mcc01b}. They use nitrate enhancements associated with events of known proton fluence (e$.$g$.$ the 1972 and 1989 flares) to determine a scale factor between fluence and nitrate enhancement.  This allows an estimate of fluence given a measured nitrate enhancement, with a range based on possible scale factor variation.

We assume a fluence of protons with energies greater than 30 MeV of $27.4 \times 10^9~ \mathrm{cm^{-2}}$ for the 1859 event, corresponding to the middle of the range of estimated values in \citet{mcc01a,mcc01b}.  Given the known fluence of the October 1989 event ($4.2 \times 10^9~ \mathrm{cm^{-2}}$ \citep{mcc01a}) the 1859 event was 6.5 times more energetic in protons.  We use this value to scale up the computed atmospheric ionization profiles that were used by \citet{jack95} to study effects of the October 1989 event for use in this study.  This scaling is, of course, uncertain, since there is no way to know the specific proton spectrum for the 1859 event since this sort of data was not available before about 1955 \citep{sv75}, but it is a ``best guess'' approach.  A linear scaling seems appropriate, given that it has been shown for photon events of large fluence (which have similar atmospheric effects) that the production of nitrogen oxides ($\mathrm{NO_y}$) scales linearly with fluence, and the deposition of nitrate is directly dependent upon the $\mathrm{NO_y}$ production \citep{ejz07,thom05}.  While x-rays from the flare would be important in the upper atmosphere (above about 70 km), they do not penetrate to the stratosphere and so have little or no impact on ozone \citep{bar99,hin65}.  We have therefore neglected any effects of x-rays.

The scaled-up ionization profiles are input to the short term version of the atmospheric model as a source of $\mathrm{NO_y}$ and $\mathrm{HO_x}$ which then go on to deplete ozone through catalytic cycles \citep{jm04}.  
Several previous SPE studies have found that the proton flux is restricted to latitudes above about $60^\circ$ \citep{mcp81,jack01,jack05b} by the Earth's magnetic field, the structure of which is modified by the SPE.  We have no way of knowing the precise latitude restriction of the proton flux in the 1859 case, but we adopt the previous limit as likely.

We scale the whole range of ionization rates from the October 1989 event by a factor of 6.5, including ionizations that result from protons with energies between 1 and 30 MeV.  The $\mathrm{HO_x}$ constituents, which are especially important above 50 km, are greatly impacted but have relatively short lifetimes and their effect is gone within several hours after the event is over.  $\mathrm{NO_y}$ constituents above 50 km are increased by these lower energy protons and can be transported downwards to the upper stratosphere (below 50 km) during late fall and winter \citep{jack05b}.  The largest stratospheric impact will be by those protons with energies greater than 30 MeV, because the $\mathrm{NO_y}$ produced by these high energy protons will be much deeper into the stratosphere where the lifetime of the $\mathrm{NO_y}$ family can be quite long (months in the middle stratosphere to years in the lower stratosphere).

While the main magnetic storm associated with the 1859 flare was observed to last about 2 hours, the particle event likely occurred over several days, perhaps as many as 10 days.  Most SPEs have durations of several days; the 1989 SPE from which we are extrapolating lasted 12 days.  We have chosen to input ionization from the 1859 SPE over 2 days in our model.  This duration may be shorter than the actual event, though we note that the magnetic storm had a short duration.  This is a convenient duration for practical purposes with the model we used.  It is known that long term atmospheric effects (e$.$g$.$ ozone depletion) from such ionization are much more strongly dependent on total fluence than on duration \citep{ejz07}.  Therefore, we believe a difference in duration is not likely to yield significant changes to our conclusions.  An expanded study could check this assertion in the context of proton events.

The total ionization is distributed over the 2 day duration uniformly (i$.$e$.$ as a step function) in the middle of a 7 day run of the short term model.  Results of this run are then read in to the long term model which is run for several years to return the atmosphere to equilibrium, pre-flare conditions.
 
\section{Results}
\label{sec:results}
\begin{figure}
\noindent\includegraphics[width=20pc]{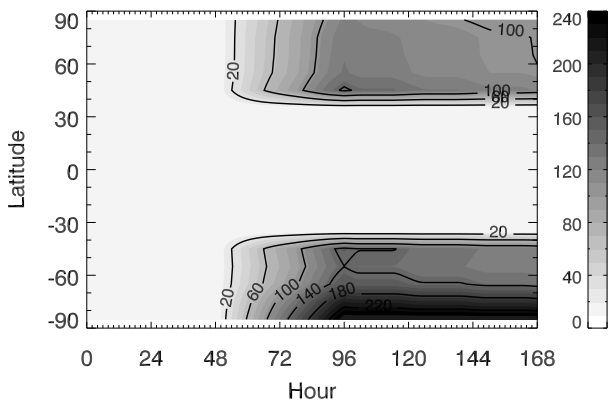}
\caption{Percent difference in column $\mathrm{NO_y}$ between perturbed and unperturbed runs during the week in which the SPE occurs.
\label{fig:NOy-perchg-shortterm}}
\end{figure}

Our primary results are changes in $\mathrm{NO_y}$ and $\mathrm{O_3}$ in the stratosphere.
$\mathrm{NO_y}$ is produced in the high latitude areas where the protons enter the atmosphere.  Figure~\ref{fig:NOy-perchg-shortterm} shows the $\mathrm{NO_y}$ generated during and shortly after the event, as the percent difference in column $\mathrm{NO_y}$ between a run with the effect of the SPE included and one without.  The maximum localized increase in column $\mathrm{NO_y}$ is about 240\%. 

\begin{figure}
\noindent\includegraphics[width=20pc]{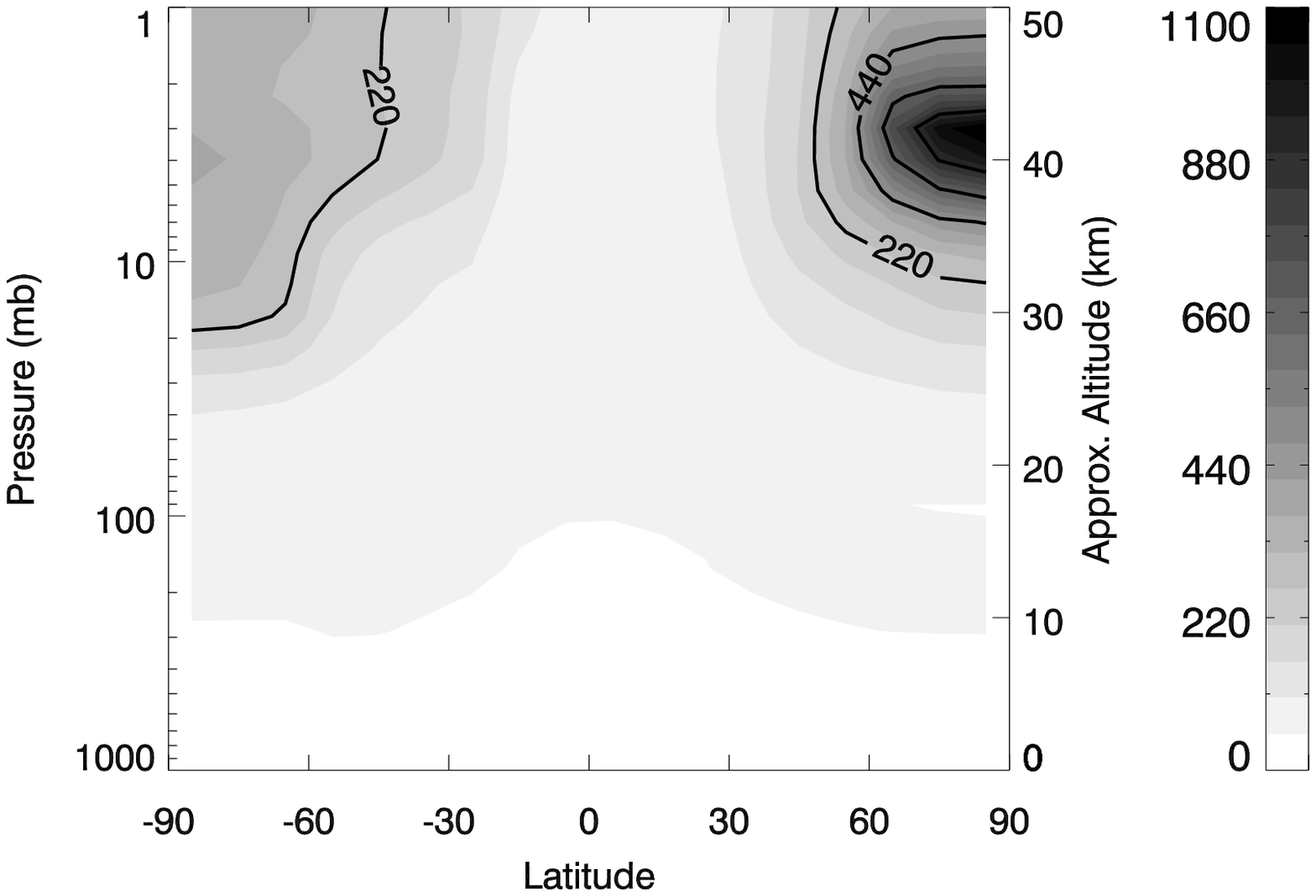}
\caption{Percent difference in profile $\mathrm{NO_y}$ between perturbed and unperturbed run, at two months after the event.
\label{fig:NOy-altlat}}
\end{figure}

\begin{figure}
\noindent\includegraphics[width=20pc]{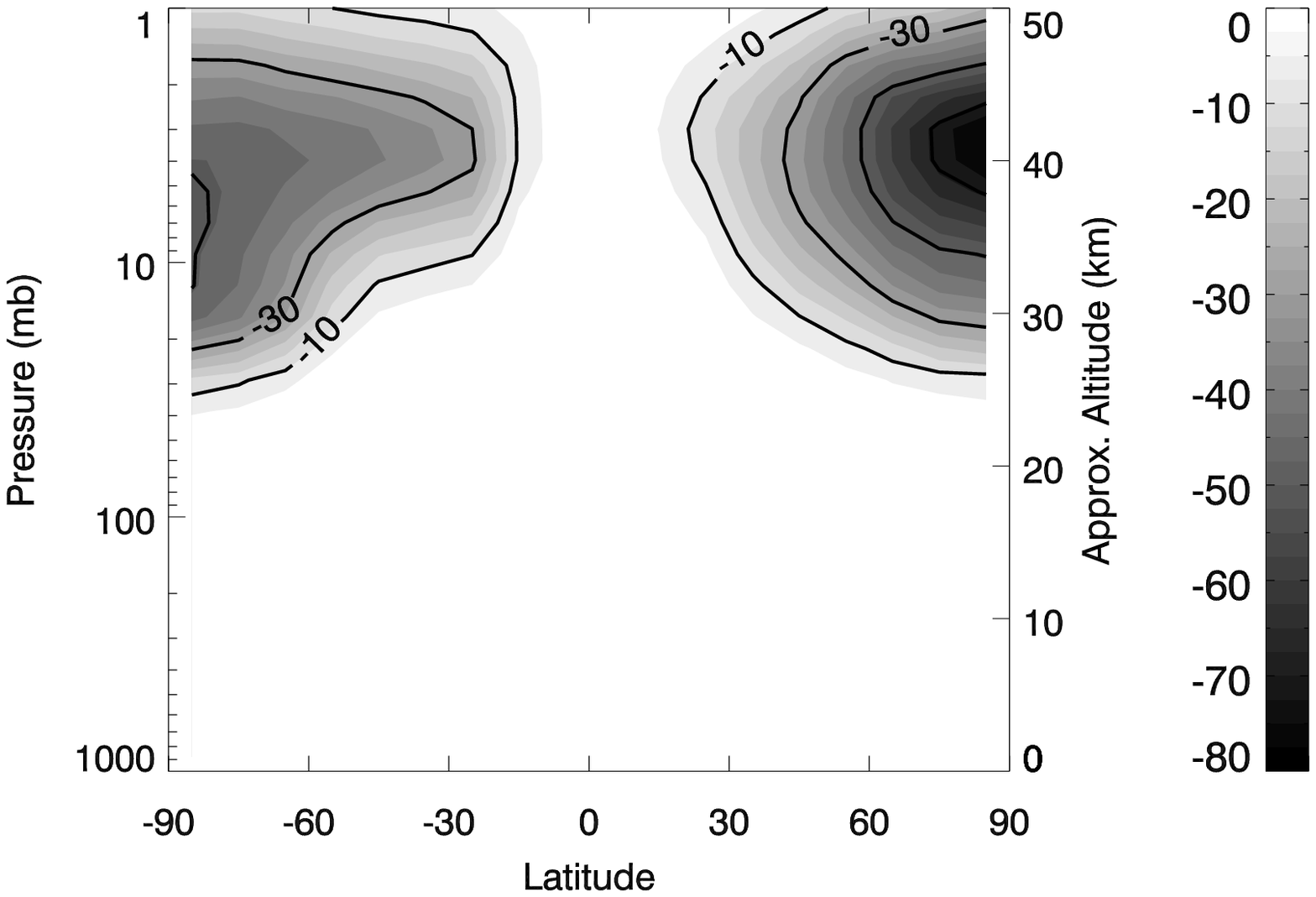}
\caption{Percent difference in profile $\mathrm{O_3}$ between perturbed and unperturbed run, at two months after the event.
\label{fig:O3-altlat}}
\end{figure}

Figures~\ref{fig:NOy-altlat} and~\ref{fig:O3-altlat} show the percent difference in profile $\mathrm{NO_y}$ and $\mathrm{O_3}$, respectively, between a run with the effect of the SPE included and one without, as a function of altitude and latitude, two months after the event.  This is the point in time when the globally averaged ozone depletion is largest (see Figure~\ref{fig:O3-glob-perchg}).  Note that the increased $\mathrm{NO_y}$ extends primarily upward in altitude from about 30 km and is most widespread in altitude at latitudes above about $30^{\circ}$.  Also, the $\mathrm{O_3}$ change is contained primarily within a band around 40 km altitude and restricted to latitudes above $30^{\circ}$.  

\begin{figure}
\noindent\includegraphics[width=20pc]{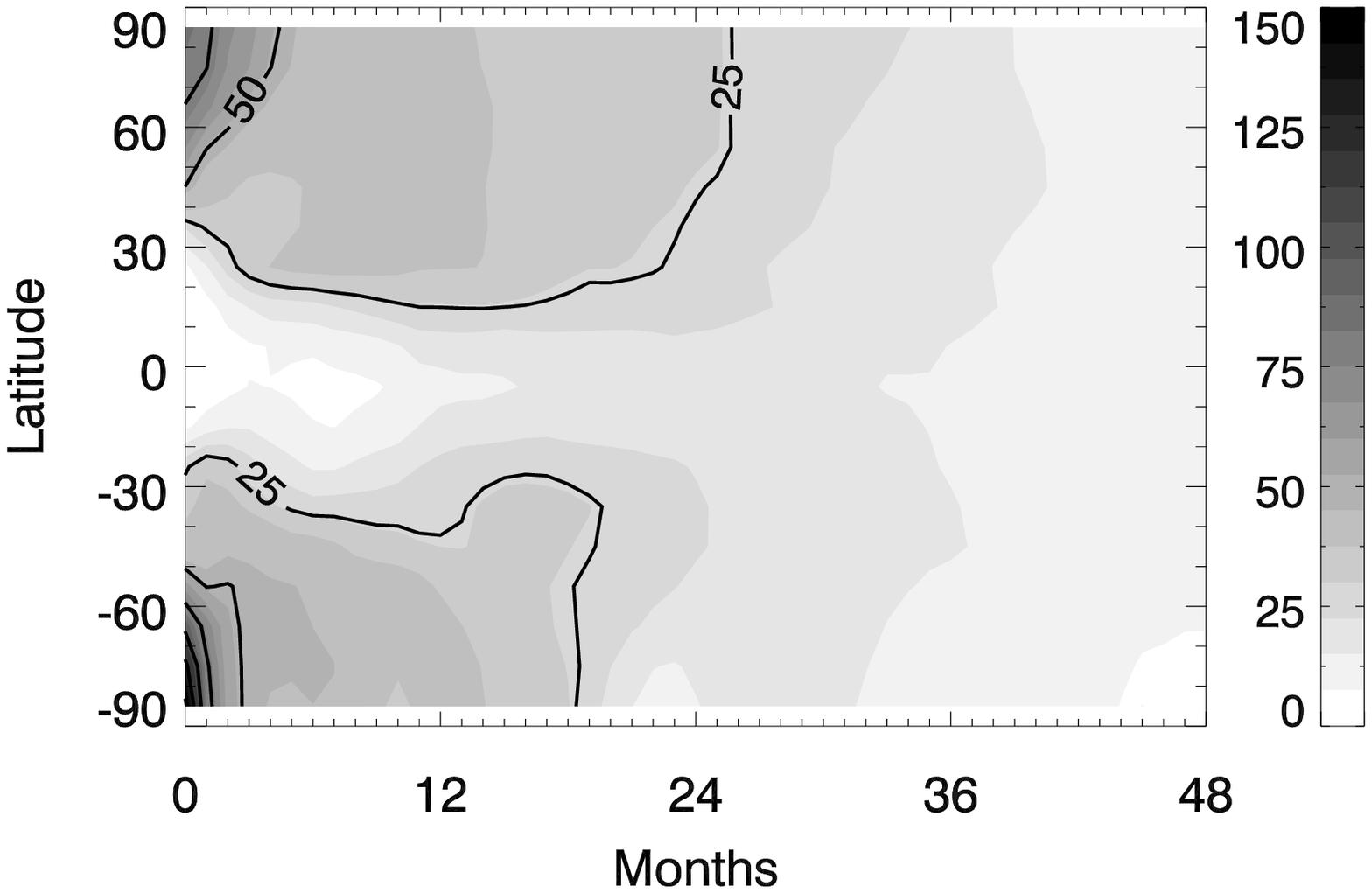}
\caption{Percent difference in column $\mathrm{NO_y}$ between perturbed and unperturbed runs for the first four years after the SPE.
\label{fig:NOy-perchg-longterm}}
\end{figure}

Figure~\ref{fig:NOy-perchg-longterm} shows the percent difference in column $\mathrm{NO_y}$ between a run with the effect of the SPE included and one without for four years after the event.  As is apparent from this plot, $\mathrm{NO_y}$ is transported to some degree to mid and low latitudes, but remains primarily concentrated in the high latitude regions as the atmosphere recovers to its pre-event equilibrium.

Figure~\ref{fig:O3-perchg} shows percent difference in column $\mathrm{O_3}$ between a run with the effect of the SPE included and one without for four years after the event.  The maximum localized decrease in $\mathrm{O_3}$ column density is about 14\% and occurs in the high latitude areas where the $\mathrm{NO_y}$ increase is largest.  

One may notice in Figures~\ref{fig:NOy-perchg-shortterm}-\ref{fig:O3-perchg} that there is asymmetry between northern and southern hemisphere regions.  This is because levels of $\mathrm{NO_y}$  and  $\mathrm{O_3}$ vary seasonally, especially in the polar regions, due to variations in the presence and intensity of sunlight.  Photolysis reactions play a critical role in the balance of constituents here and strongly affect the total values.  A detailed discussion of these effects can be found in \citet{thom05} for the case of a gamma-ray burst.

Figure~\ref{fig:O3-glob-perchg} shows the globally averaged percent difference in ozone.  The maximum decrease is about 5\%, which occurs two months after the event.  Maximum global ozone depletion is delayed as $\mathrm{NO_y}$ spreads and interacts with $\mathrm{O_3}$ over a larger area.  This is quantitatively similar to the globally averaged anthropogenically-caused ozone depletion currently observed, which is predicted to diminish slowly over several decades \citep{wmo03}.  By contrast, this naturally-caused ozone depletion from the 1859 solar flare is nearly gone in about four years.

Since we have estimated the intensity of the 1859 SPE using data from nitrate deposition in ice cores, a consistency check may be done by computing the nitrate rainout in the model.  We have approached such a comparison in two ways.  

First, the maximum localized enhancement over background in the model is about 14\%, while the 1859 spike in Figure 1 of \citet{mcc01a} is about 200\% over background.  This is obviously a large disagreement.  However, because nitrate deposition can be spotty and takes place over a period of months, we have also computed absolute deposition values.  Adding up deposition in the model over three months following the flare, within the $10^{\circ}$ latitude band centered at $75^{\circ}$ North, yields a value of $1360 \hspace{1mm}\mathrm{ng \hspace{1mm}cm^{-2}}$.  A similar computation following \citet{mcc01a,mcc01b}, using the fluence value assumed above and a value of 30 for the conversion factor between fluence and nitrate deposition \citep{mcc01a}, gives $822 \hspace{1mm}\mathrm{ng \hspace{1mm}cm^{-2}}$, a factor of about 1.7 smaller than the model value.  

We note that the difference between modeled and observed absolute deposition values is opposite that of the percent enhancement over background.  That is, the percent enhancment in the model is smaller than that from the ice core data, while absolute values from the ice core data are smaller than in the model.  Given the many sources of uncertainty the absolute comparison is at least reasonably close, being less than a factor of two different.  

It is important to note that the percent enhancement comparison looks only at the height of the peak above background, while the absolute deposition value is effectively the area under that peak.  A difference in how the nitrate is deposited over time could help explain this discrepancy.  If in the actual event most of the nitrate was deposited in a relatively short amount of time compared to the model then the height of the peak would be greater, even if the area under that peak is less, as seen in our comparisons.

Similar comparisons have been done before with this model \citep{jack90,thom05}.  \citet{jack90} found that the model showed a smaller peak percentage enhancement of nitrates than did ice core results of \citet{zel86} (around 10\% from the model, compared to 400\% from ice core).  This discrepancy is of the same order as what we have found in this study and apparently indicates some scaling disagreement between the model computations and the observations.

\section{Conclusions}
\begin{figure}
\noindent\includegraphics[width=20pc]{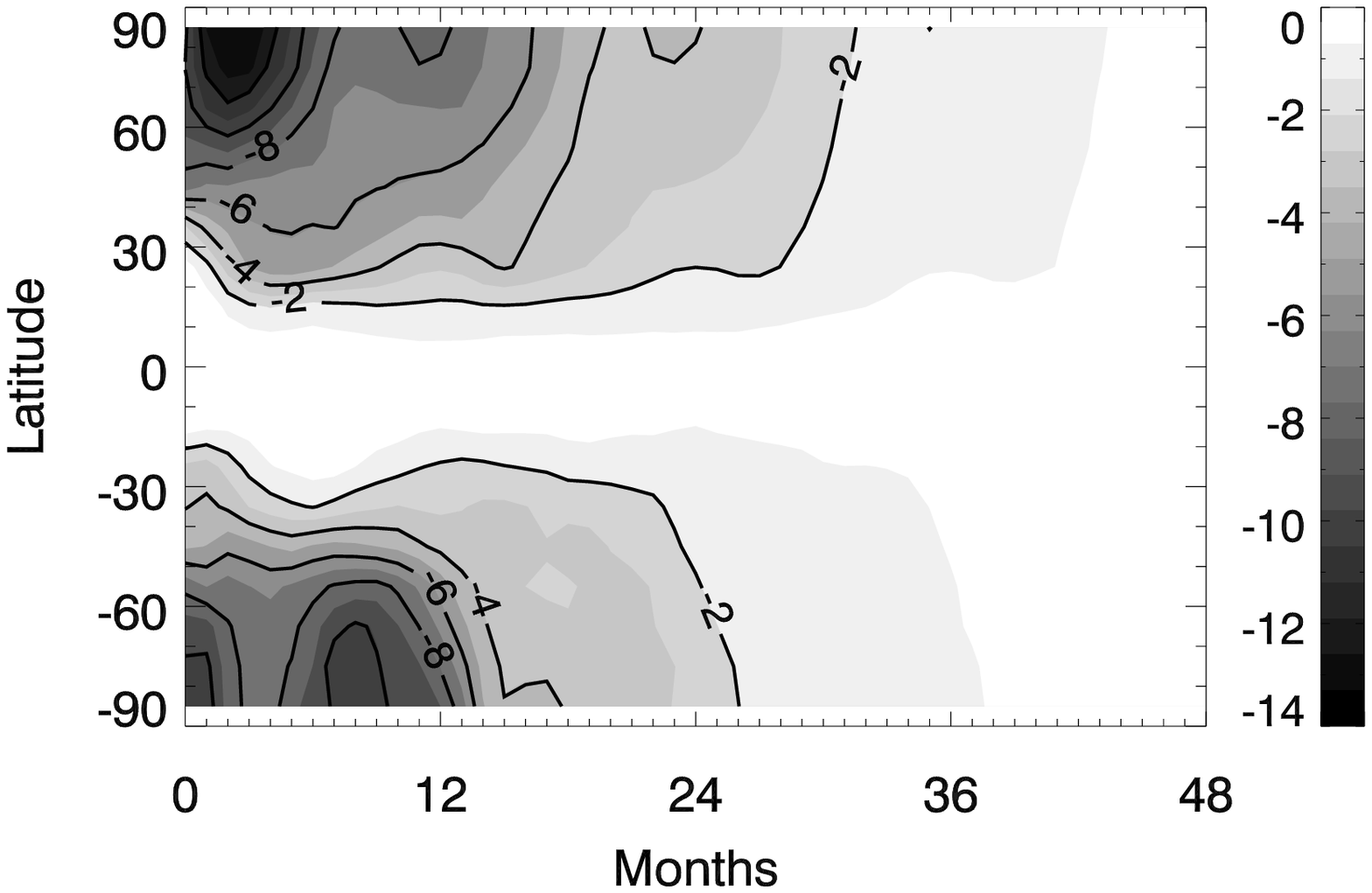}
\caption{Percent difference in column $\mathrm{O_3}$ between perturbed and unperturbed runs for the first four years after the SPE.
\label{fig:O3-perchg}}
\end{figure}

\begin{figure}
\noindent\includegraphics[width=20pc]{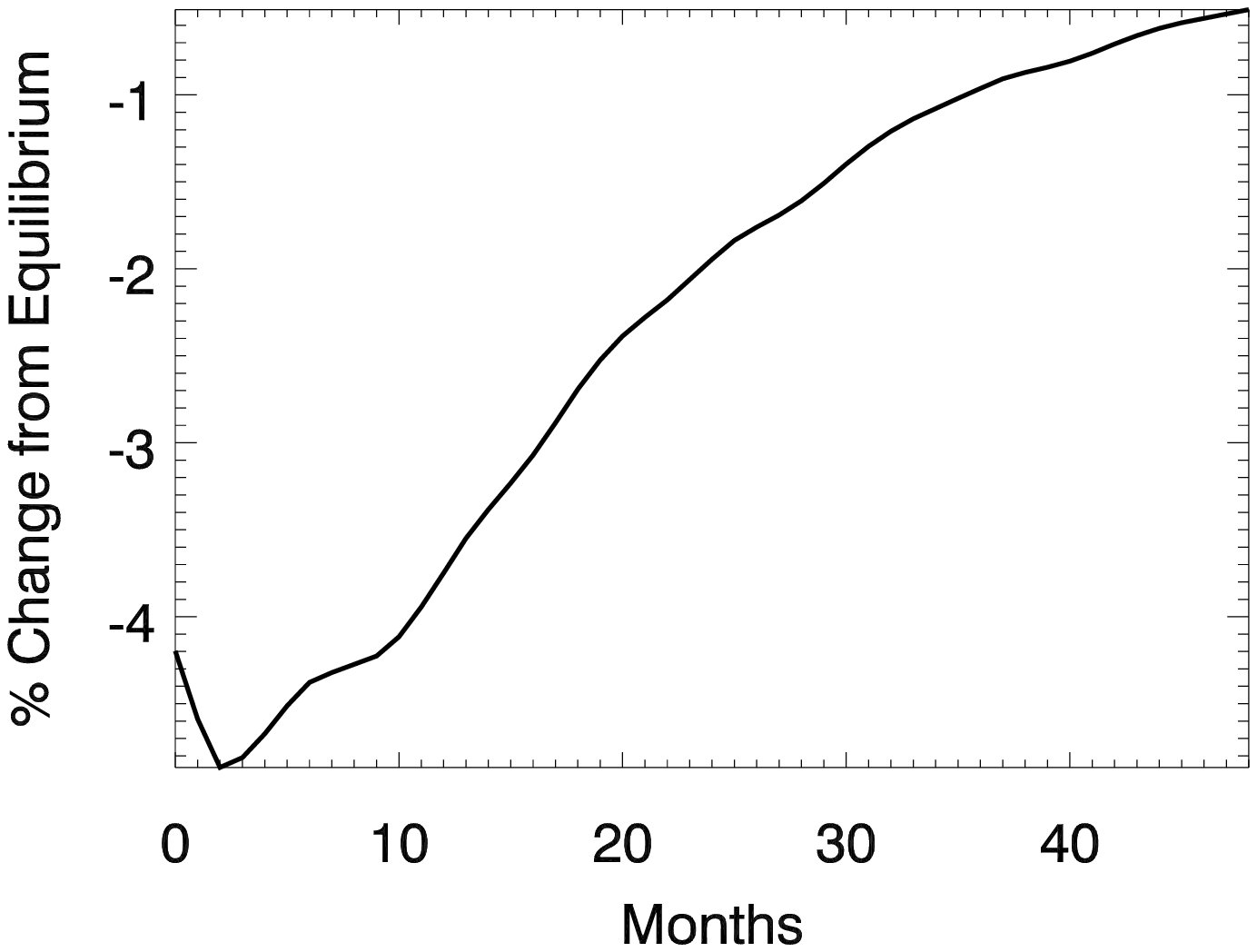}
\caption{Gobally averaged percent difference in column $\mathrm{O_3}$ between perturbed and unperturbed runs for the first four years after the SPE.
\label{fig:O3-glob-perchg}}
\end{figure}

We find for the 1859 event an atmospheric impact appreciably larger than that of the most energetic flare in the era of satellite monitoring, that of October 1989.  Localized maximum column ozone depletion (see Figure~\ref{fig:O3-perchg}) is approximately 3.5 times greater than that of the 1989 event (see Figure 3 of \citet{jack95}).  We note that this is a smaller factor than that relating the total energy of the two events (6.5).  Ozone depletion has been seen in other contexts (e.g., gamma-ray burst impacts) to scale less than linearly with total energy \citep{thom05,ejz07}.  
Causes for this weaker dependence include: increased removal of important depletion species such as NO with increased levels of $\mathrm{NO_y}$; production of $\mathrm{O_3}$ in isolated regions especially at lower altitudes which are normally shielded from solar UV which produces  $\mathrm{O_3}$; and the ``saturation'' of depletion, where most of the  $\mathrm{O_3}$ in a given region is removed and so cannot be depleted any further.

Our nitrate deposition results do not show as dramatic an enhancement as the measurements by \citet{mcc01a}.  This discrepancy does mirror a similar comparison between nitrate deposition as computed using this model and that measured in ice cores as described in \citet{jack90}.  This discrepancy may reflect differences in transport or deposition efficiency.

Finally, while the ozone depletion seen here is limited, even small increases in UVB can be detrimental to many life forms \citep{rou99,cull92}.  Flares of significantly larger energy have been observed on Sun-like stars \citep{sch00}, and may occur from time to time through the long history of life on Earth. (See \citet{hw03} for an overview of terrestrial extinctions.)  Such events would have more dramatic effects on the biosphere.  Knowledge of the impacts of such flares is important in understanding the history of life here and possibly elsewhere, in particular on terrestrial planets around stars which are more active than the Sun.

\begin{acknowledgments}
This work was supported in part by the NASA Astrobiology: Exobiology and Evolutionary Biology Program under grant number NNG04GM14G at the University of Kansas.
\end{acknowledgments}

\end{article}

\end{document}